# TITLE: Pressure-Controlled Layer-by-Layer Oxidation of $ZrS_2$(001) Surface

*AUTHOR NAMES*

Liqiu Yang[1], Rafael Jaramillo[2], Rajiv K. Kalia[1], Aiichiro Nakano[1*], and Priya Vashishta[1]

AUTHOR ADDRESS

[1]*Collaboratory for Advanced Computing and Simulation, University of Southern California, Los Angeles, CA 90089-0242, USA*

[2]*Department of Materials Science and Engineering, Massachusetts Institute of Technology, Cambridge, MA 02139, USA*

**Corresponding Author**

*Corresponding author. Email: anakano@usc.edu

ABSTRACT

Understanding oxidation mechanisms of layered semiconducting transition-metal dichalcogenide (TMDC) is important not only for controlling native oxide formation but also for synthesis of oxide and oxysulfide products. Here, reactive molecular dynamics simulations show that oxygen partial pressure controls not only the $ZrS_2$ oxidation rate but also the oxide morphology and quality. We find a transition from layer-by-layer oxidation to amorphous-oxide-mediated continuous oxidation as the oxidation progresses, where different pressures selectively expose different oxidation stages within a given time window. While the kinetics of the fast continuous oxidation stage is well described by the conventional Deal-Grove model, the layer-by-layer oxidation stage is dictated by reactive bond-switching mechanisms. This work provides atomistic details and a potential foundation for rational pressure-controlled oxidation of broad TMDC materials.

TOC

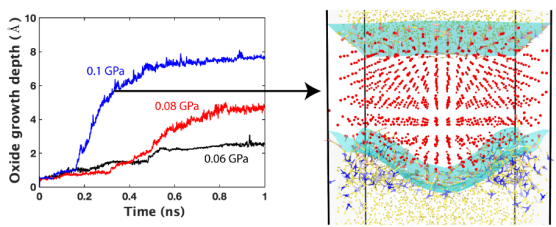

TEXT

Transition-metal dichalcogenides (TMDC) are promising materials for their potential applications in next-generation electronic and optoelectronic devices. Among them, semiconducting $ZrS_2$ [1,2] is especially suitable for microelectronics [3-5]. In addition, $ZrS_2$ with the bandgap of 1.7 eV has absorption peaks in the visible-light range (400-760 nm), thus has been studied for solar-energy applications [4, 6-8]. However, native oxidation remains a major issue for many TMDCs in achieving their long-term stability, and $ZrS_2$ is no exception [9, 10]. To control the native oxidation, it is essential to understand atomistic oxidation mechanisms of $ZrS_2$. Furthermore, oxidation products themselves are of great scientific and technological interest. $ZrO_2$ nanocrystals show chemical inertness, excellent thermal stability, high hardness [11, 12], and excellent optical properties, which make them excellent materials for fuel cells [13, 14], catalysis [15-17], sensors [18, 19], bioseparation [20], chromatography [21], and high-refractive-index nanocomposites [22]. Tetragonal zirconium oxysulfide (ZrOS) nanopowder synthesized by sol-gel method was proposed for thin film photovoltaic applications and dielectric material [23]. In the first-principles study by Zhang *et al*. [24], Janus monolayer ZrOS was found to have large dielectric permittivity, mechanical and dynamical stabilities, high and sharp absorption peaks in visible and ultraviolet (UV) light range, showing great potential for dielectric semiconductor material and visible and UV light sensor applications.

Oxidation mechanisms of $ZrS_2$ have been studied under various conditions. Li *et al*. [25] studied the oxidation of monolayer $ZrS_2$ and found that surface vacancies and edges have high affinity towards the adsorption and activation of oxygen molecules. In a combined experiment and simulation study, Jo *et al*. [9] observed a higher oxidation rate of $ZrS_2$ than that of $MoS_2$, and identified the key atomistic mechanisms: rapid $O_2$ adsorption and bond scission, followed by oxygen transport into the crystal *via* Zr-O bond switching and the collapse of van der Waals (vdW) gaps, resulting in the formation of an intermediate amorphous oxy-sulfide. It was also suggested that the oxidation is diffusion-controlled in the fast oxidization period [9, 26]. Using reactive molecular dynamics (RMD) simulations with an optimized reactive force field, Yang *et al*. [26] found a higher oxidation rate on (210) surface compared to (001) surface, where the oxidation in the fast oxidization period on both surfaces were identified as diffusion-controlled. In humid environment, it was suggested that $H_2O$ contributes to the breakdown of $ZrS_2$ not only because a $H_2O$ molecule can adsorb on the metal to achieve continuous oxidation, but also because its strong polarization promotes the adsorption of nonpolar oxygen molecules [25, 27].

An outstanding issue is oxidation behavior in different environments[28], which is important for device processing and technology development. Here, we use RMD simulations to investigate how oxidation of the ZrS$_2$(001) surface depends on oxygen pressure. The oxidation rate increases under high O$_2$ partial pressure. We find that, as oxidation progresses, there is a transition from layer-by-layer oxidation to amorphous-oxide-mediated continuous oxidation. The first stage, layer-by-layer oxidation, proceeds with reactive bond switching and rotation events, leading to the closure of vdW gaps. In the second stage, the kinetics of continuous oxidation and its pressure dependence are well-described by the conventional Deal-Grove model [29]. Different pressures can selectively expose different stages of oxide growth within a given time window—from layer-by-layer *via* reaction-limited linear to diffusion-limited parabolic growth—with increased pressure. We also find a trade-off between the growth rate and the quality of the grown oxide structure, as well as pressure control of the morphology of semiconductor/oxide interfaces. Overall, our findings reveal excellent pressure-controllability of the oxidation of layered vdW semiconductors.

**Results**

We performed RMD simulations to study the oxidation of a ZrS$_2$(001) slab in oxygen (O$_2$) atmosphere with varying pressure at temperature 300 K (Figure 1, a and b).

***Oxide growth dynamics*:** Figure 1, c-e, shows snapshots of the system after 0.6 ns of oxidation at oxygen pressures of 0.1, 0.08 and 0.06 GPa. Oxidation proceeds the fastest at the highest pressure of 0.1 GPa, followed by that under 0.08 GPa, and slowest at 0.06 GPa.

To better represent the pressure dependence of the oxidation rate, Figure 2 shows snapshots of ZrS$_2$ oxidation at various pressures and times, in which oxidation fronts are shown as surfaces and sulfur (S) atoms are omitted for clarity. Oxygen (O) atoms are colored yellow, and zirconium (Zr) atoms are colored according to the coordination number to O atoms. At time 0.2 ns, Figure 2, a-c, shows a similar initial oxidation stage for all pressures. The pressure dependence becomes evident by time 0.6 ns. At the highest pressure of 0.1 GPa, oxidation of the second layer is completed by 0.6 ns, and the oxidation front has reached the third layer (Figure 2d). In contrast, the oxidation front has just reached the second layer at 0.08 GPa (Figure 2e) and is still in the first layer at 0.06 GPa (Figure 2f).

To quantify the pressure dependence of the oxidation rate, Figure 3a shows the oxide depth as a function of time. We define the oxide depth as the average position of oxygen atoms in the solid, taking the origin as the $ZrS_2$ surface. At 0.1 GPa (blue curve in Figure 3a), oxide growth begins slowly, accelerates around 0.16 ns, and then gradually slows down, thus exhibiting a two-stage oxidation behavior. Fast oxidation in the second stage is characterized by a parabolic shape, indicative of a diffusion-controlled process, as seen in our previous work [26]. At a lower pressure of 0.08 GPa (red curve in Figure 3a), we observe similar behavior: oxidation is slow at first, then becomes faster around 0.3 ns. At the lowest pressure of 0.06 GPa, oxide growth remains slow throughout. The slow oxidation process at 0.06 GPa is characterized by stepwise increase of oxide depth, indicative of layer-by-layer ooxidization, as found in our previous work [26]. Here, oxidation of each layer begins with a kink, followed by flattening of the oxide front with the completion of the layer's oxidation, resulting in the observed stepwise increase in oxide depth. The oxidation process exhibits similar layer-by-layer growth kinetics at short times (*i.e.*, before the onset of fast growth) for all pressures. Similar kink mechanism (*i.e.*, nucleation of a spatially-localized growth front, followed by lateral growth to form a flat surface) is characterized by a low activation energy, thus observed in many processes such as domain-wall motion [30] and crack extension [31]. It appears that, after a period of layer-by-layer oxidation, a transition occurs to amorphous-oxide-mediated continuous oxidation, with the transition occurring at earlier times with increasing pressure.

To probe chemical reactions underlying the crossover of oxidation stages, Figures. 3b and 3c show time evolution of the numbers of Zr-O and Zr-S bonds, respectively. Bond order is calculated for all Zr-O and Zr-S bonds, and bond is counted only when bond order is larger than 0.3. Figure 3b shows increase in the number of Zr-O bonds as the oxidation progresses. The increase is slowest at the lowest pressure of 0.06 GPa. At the highest pressure of 0.1 GPa, there is an abrupt increase in the number of Zr-O bonds at 0.16 ns, which is consistent with the onset of the fast oxidation in Figure 3a. A notable observation is that the number of Zr-O bonds increases faster at 0.08 GPa than at 0.1 GPa. This could be due to disruptive and incomplete oxidation within short time at the highest pressure of 0.1 GPa, which in turn produces low-quality defective oxide. To quantify this effect, Figure S1 in supplementary information shows Zr-O partial pair distribution at various times for all pressures. At time $t \leq 0.2$ ns, the first peak is located at 1.75 Å indicative of Zr-O chemical bond at all pressures. After 0.3 ns under the highest pressure of 0.1 GPa, however, we observe the lowering of the first peak at 1.75 Å accompanied by the development of large correlation at distance larger than 2 Å, signifying mechanical disordering of chemical structures. This explains the decrease in the Zr-O bond number after 0.3 ns (Figure 3b), while the oxide is still growing (Figure 3a); note that Zr-O pairs farther than 2 Å are not chemically bonded nor counted in Figure 3b. In fact, Figure 2d shows some surface Zr atoms to be undercoordinated (colored light blue). These surface Zr atoms are not fully coordinated with oxygen atoms, and thus oxygen atoms are able to move faster inward without being properly coordinated with Zr. Figure 3c shows a decrease in the number of Zr-S bonds as a function of time, where the decrease is fastest at the highest pressure of 0.1 GPa. Again, there is an abrupt change of the rate of chemical reactions at 0.16 ns for the 0.1 GPa case, which is consistent with the observations in Figures 3a and 3b. In supplementary information, Figure S2 shows local structures at 0.6 ns under 0.1 GPa, whereas Figures S3 and S4 compare Zr-O bond-length and O-Zr-O bond-angle distributions, respectively, at 0.6 ns under various pressures.

To elucidate the different stages of oxide growth, Figure 4 shows the oxidation front at different times for the 0.1 GPa simulation. At time $t = 0.2$ ns, oxidation is confined in the first layer shown in Figure 2a. By 0.3 ns (Figure 4a), the oxidation front has reached the second layer, and by 0.4 ns it has reached the third layer (Figure 4b). The oxide surface at this time is not flat, showing kinks in the middle. At time $t = 0.5$ ns (Figure 4c), the oxide surface has become flatter again, due to oxidation of $ZrS_2$ in the middle of the frame. This suggests that a layer-by-layer like oxidation process continues even during the second stage of faster oxidation, but the oxidation front now encompasses multiple vdW layers. As a result, the average oxide depth increases continuously (Figure 3a) instead of stepwise despite the underlying kink-mediated layer-by-layer oxidation. A similar layer-by-layer oxidation process was reported in $Cu_2O$ nano-island growth [32] and twin-boundary-assisted oxidation in Ag nanocrystals [33].

**Discussion**

*Applicability of the Deal-Grove model*: Oxide growth behavior after the sudden onset of fast oxidation is characterized by gradually decreasing oxidization rate similar to the Deal-Grove model of silicon thermal oxidation, *i.e.*, linear growth controlled by interfacial chemical reaction followed by parabolic growth indicating a diffusion-controlled process [29] (see Figures S5 and S6, along with associated discussion in supplementary information). Oxidation of $ZrS_2$ in the parabolic-growth stage has recently been proposed to be diffusion-controlled [9, 26]. Figure 3a exhibits both parabolic (at 0.1 GPa) and linear (at 0.08 GPa) oxide growth in the second oxidation stage. Such linear-to-parabolic oxidation is conventionally described by the Deal-Grove model [29]. The model assumes that oxidation proceeds by inward movement of oxidant species rather than by outward movement of reactants. The process applies after an initial transient period (see discussion below) with the consequence that the fluxes of oxidant in each of the steps are identical at all times. Key assumptions for the Deal-Grove model are: (1) the oxidants are transported from the oxidizing gas to the outer surface, where they react or are adsorbed; (2) they are then transported across the oxide film towards the bulk; (3) they finally react at the oxidation front to form a new layer of oxide. In applying these assumptions to the oxidation of $ZrS_2$, we note differences that require examination. On one hand, it is important to clearly identify the reaction products, *i.e.*, whether there are sulfur atoms remaining after oxidation, and when and how $SO_2$ are emitted if the final products include $SO_2$. During our simulation timespan (< 0.6 ns), we did not observe $SO_2$ formation. Accordingly, we do not consider the formation of gas products and their out-diffusion here. We can then assume that the oxidation process, A(solid) + B(gas) = C(solid), follows a first-order reaction during the simulation time window. On the other hand, there is a spatial gap between consecutive $ZrS_2$ layers unlike bulk silicon, and thus the deviation from the Deal-Grove model resulting from the gap needs to be considered. In our simulation, the lattice constant for the surface-normal direction is 5.85 Å and the gap between vdW layers is 2.925 Å, thus an effective diffusion coefficient incorporating the effect brought by the gap should be considered.

In Figure 5, we present fits of the Deal-Grove model to our simulation results for pressures 0.08 and 0.1 GPa. The Deal-Grove equation describes the oxide growth behavior at both linear and parabolic growth stages:

$$d_0^2 + A d_0 = B(t + \tau), \tag{1}$$

where $d_0$, $t$ and $\tau$ represent oxide depth, oxidation time, and shift in the time to correct for the presence of the initial oxide layer. In our case, $\tau$ can be identified as the onset time of the fast continuous oxidation in the second stage, after the slow layer-by-layer oxidation in the first stage shown in Figure 3a. We thus fit the data points using the Deal-Grove model only beyond the oxide depth of $d_0(\tau = 0.3 \text{ ns}) = 0.76724$ Å for 0.08 GPa and $d_0(\tau = 0.16 \text{ ns}) = 0.90787$ Å for 0.1 GPa. At the lowest pressure of 0.06 GPa, the finite simulation time only places a lower bound of 0.6 ns for $\tau$ (Figure 5c), before which Eq. (1) does not apply. Oxidation for $t > \tau$ is well-modeled by a linear law at 0.08 GPa (Figure 5b), while it exhibits both linear and parabolic behaviors at 0.1 GPa (Figure 5a). For long oxidation time ($t \gg \tau$ and $t \gg t_c = A^2/4B$), Eq. (1) describes a parabolic growth law, $d_0^2 \cong Bt$. The parabolic rate constant, $B$, is proportional to the partial pressure $p$ of the oxidizing species in the gas phase according to Henry's law [29]. Accordingly, the crossover time, $t_c = A^2/4B$, which separates the linear ($t < t_c$) and parabolic ($t > t_c$) growth regimes [29], is inversely proportional to $p$. At the highest pressure of 0.1 GPa, the estimated $t_c$ is 0.25 ns and the fittings in Figure 5 exhibits a linear-to-parabolic crossover, which is expected from Eq. (1). On the other hand, only the first linear regime is observed within the simulated time at a lower pressure of 0.08 GPa, indicating a much longer crossover time, $t_c > 0.6$ ns. The pressure dependence of the crossover time is thus consistent with the Deal-Grove model. The short-time behavior of Eq. (1) for $t \ll \tau$ follows a linear law, $d_0 \cong (B/A)(t + \tau)$, which is dictated by interfacial reactions [29]. From the linear fits in Figure 5, the estimated $B/A = 87.3$ and 361 cm/s, respectively at $p = 0.08$ and 0.1 GPa, which is again consistent with the pressure dependence expected from the Deal-Grove model. The parabolic regime is only probed for $p = 0.1$ GPa in our simulation data, which yields $B = 3.42 \times 10^{-6}$ cm²/s. While larger than typical diffusion constants in ambient thermal oxidation experiments, this value is an orders-of-magnitude smaller than those in high-pressure oxidation simulation of Al [34].

Overall, the Deal-Grove model provides an excellent platform to mechanistically understand essential pressure control of the three stages of oxidation observed in our simulations:

- Stage I ($t < \tau$): Discrete layer-by-layer growth.

- Stage IIa ($\tau < t < t_c$): Short-time continuous growth following linear law, which is dictated by interfacial reactions.

- Stage IIb ($t_c < t$): Long-time continuous growth following parabolic law, which is dictated by oxidant diffusion.

The two crossover times, $\tau$ and $t_c$, are both pressure-dependent and thus are controllable by oxygen partial pressure. Our three simulations selectively expose these three stages of oxide growth within a given time window: only stage I at 0.06 GPa; stages I + IIa at 0.08 GPa; and stages I + IIa + IIb at 0.1 GPa. In particular, layer-by-layer oxidation stage can be prolonged at lower pressures, which may be useful for device processing steps that require highly controlled and ultra-thin oxide formation.

*Initial oxidation*: We study pressure-dependence of the initial oxidation process before time $\tau$, by computing the change of Gibbs free energy $\Delta G$ upon oxygen-atom adsorption on the ZrS$_2$(001) surface (details of the free energy calculation are provided in supplementary information). In Figure S7, we present $\Delta G$ for oxygen-atom adsorption on Zr sites as a function of oxygen chemical potential $\Delta\mu_O$. The bars under the plot show the corresponding pressure at temperatures $T$ = 300, 600 and 900 K. The calculated $\Delta G$ is negative, even below 1 atm, which suggests that initial oxidation is driven by strongly exothermic Zr-O binding.

After the initial period of oxygen adsorption and layer-by-layer oxidation, we observe a process of more rapid oxygen diffusion and an amorphous oxide growth front that spans multiple vdW layers. We find that oxygen diffusion is intimately related to Zr-site disorder, as suggested by our previous RMD simulations [26]. In Figure 6, we illustrate how disordered Zr atoms assist oxygen motion between vdW layers, using two instances drawn from the simulation at 0.1 GPa. In each case, a Zr atom (labeled Zr1) at the oxidation front has moved into a vdW gap, where it bonds with S atoms in adjacent layers. We observe that the O atom (labeled O1) moves across the vdW gap by transient bonding with Zr1, thus facilitating oxygen diffusion in to the next ZrS$_2$ layer. We speculate that this mechanism is facilitated by the fact that sulfur is a soft ion that can adopt multiple oxidation states, and therefore can be readily oxidized and reduced by virtue of the motion of hard ions O$^{2-}$ and Zr$^{4+}$. The formation of kinks in the oxide growth front (Figure 4) may result from the local acceleration of oxygen diffusion and subsequent oxide formation by Zr defects. This amorphous-oxide-mediated oxidation, in turn, explains the rapid increase in the oxide depth and the number of Zr-O bonds in Figure 3.

**Conclusion**

Our reactive molecular dynamics simulations supported by first-principles calculations elucidate the atomistic mechanisms and pressure-dependence of oxidation of pristine $ZrS_2$(001) surfaces. The initial adsorption of oxygen atoms on the surface is driven by the large binding energy. After an initial stage of slow, layer-by-layer oxidation, local breakdown of van der Waals gaps accelerates the diffusion of oxygen, creating a kink-mediated amorphous oxide growth front, with kinetics well described by the conventional Deal-Grove model. The transition time from layer-by-layer to continuous oxidation, as well as that between reaction-limited linear and diffusion-limited parabolic oxidation within the Deal-Grove regime, is well controlled by pressure. At even longer time scales (beyond our simulation window), we expect that the kinetics may deviate from the Deal-Grove model due to the out-diffusion of $SO_2$ byproducts, which is a topic of future study. To synthesize oxide-semiconductor heterostructures in emerging two-dimensional (2D) electronics, active control of oxidation is being vigorously explored, including plasma-, ultraviolet- and laser-assisted oxidation processing [35, 36]. In such advanced growth chambers, active control of oxygen partial pressure can be used as another growth-control parameter. The current work suggests that oxygen partial pressure is indeed a promising control parameter for such active oxidation for future 2D electronics.

**Methods**

**Reactive molecular dynamics (RMD) simulations**

RMD simulations in the canonical (NVT) ensemble were performed using the Large-scale Atomic/Molecular Massively Parallel Simulator (LAMMPS) software [37]. While NVT simulation exhibits the same sequence of oxidation events as in the isothermal-isobaric (NPT) simulation (Figure S8 in supplementary information), the former corresponds more closely to advanced active-oxidation chambers that operate in a batch-reaction mode [35, 36]. The ReaxFF reactive force field was developed and optimized for ZrS$_2$ oxidation, and the details can be found in previous papers [9, 26]. An 8-layer ZrS$_2$(001) slab with 1,536 Zr atoms and 3,072 S atoms was placed in the middle of the simulation box, where the Cartesian $z$-axis is normal to the slab. O$_2$ molecules were placed above and below the slab to form gas atmosphere, where the number of O$_2$ molecule was calculated to achieve the desired pressure. Both ZrS$_2$ and O$_2$ were heated and fully relaxed at temperature of 300 K. The system temperature was controlled using Nosé-Hoover thermostat [38, 39]. Periodic boundary conditions were applied in all directions.

**Density functional theory (DFT) calculations of the Gibbs free energy change of adsorption**

The Gibbs free energy change due to adsorption is

$$\Delta G(T,p) = \frac{1}{A}(G_{O/Zr}^{slab} - G_{Zr}^{slab} - \Delta N_{Zr}\mu_{Zr} - N_O\mu_O), \quad (2)$$

where $G_{O/Zr}^{slab}$ and $G_{Zr}^{slab}$ are the Gibbs free energies of the oxygen-adsorbed surface and clean surface, respectively. $\mu_{Zr}$ and $\mu_O$ are the chemical potentials of Zr and O atoms. $A$ is the area of the surface. The chemical potential of oxygen, $\mu_O$, and its dependence on pressure, $p$, and temperature, $T$, are provided in the supplementary information. The Gibbs free energy change due to adsorption is simplified as:

$$\Delta G(\Delta\mu_O) = \frac{1}{A}(N_O E_{O/ZrS_2}^b - \Delta N_{Zr}\mu_{Zr} - N_O\Delta\mu_O), \quad (3)$$

where $E_{O/ZrS_2}^b$ is the binding energy of oxygen atoms on ZrS$_2$(001) surface. We used density functional theory (DFT) in the Vienna Ab initio Simulation Package (VASP) software [40, 41] to calculate $E_{O/ZrS_2}^b$. ZrS$_2$ was modeled using a supercell, where a five-layered ZrS$_2$ slab is constructed with a vacuum region of 25 Å to prevent interaction between periodic images. The oxygen ad-layer structures were modeled using 2 × 2 surface unit cells. We let oxygen atoms be adsorbed on both sides of the slab for a higher accuracy to eliminate asymmetry. The oxygen atoms and all atoms in the outer two ZrS$_2$ substrate layers were fully relaxed, whereas the central three layers were fixed in their bulk positions. We used the projector augmented wave (PAW) method [42] and Perdew-Burke-Ernzerhof (PBE) functional [43]. The energy cut-off for plane-wave expansion was 520 eV. The total energy and forces were converged to within 0.1 $\mu$eV per atom and 1 meV/Å, respectively. Brillouin zone was sampled over 5 × 5 × 1 Monkhorst-Pack k-point meshes [44]. The average binding energy per oxygen atom adsorbed on the surface, $E^b_{O/ZrS_2}$, was calculated as

$$E^b_{O/ZrS_2} = -\frac{1}{N_O}(E^{slab}_{O/Zr} - E^{slab}_{Zr} - \Delta N_{Zr} E_{Zr} - \frac{N_O}{2} E_{O_2}), \qquad (4)$$

where $N_O$ and $\Delta N_{Zr}$ are the number of oxygen atoms and the change in number of zirconium atoms, while $E^{slab}_{O/Zr}$, $E^{slab}_{Zr}$, $E_{Zr}$ and $E_{O_2}$ are the energy of the adsorbate-substrate system, the clean surface, the zirconium atoms, and the free oxygen molecule.

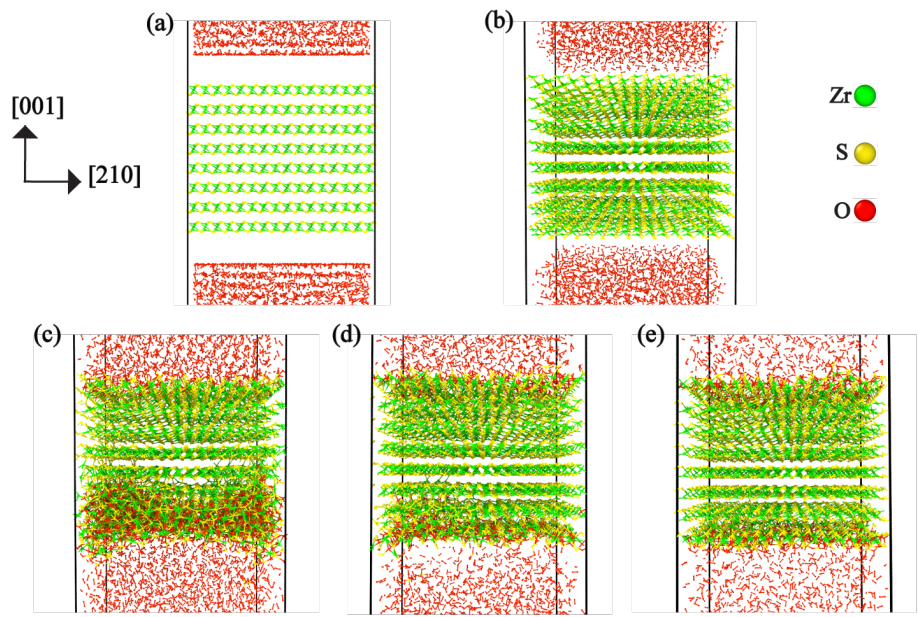

**Figure 1.** Initial and final configuration of RMD simulations of ZrS$_2$ oxidation. Initial configuration in the parallel front view (**a**) and perspective side view (**b**). Final configurations of ZrS$_2$ oxidation at time 0.6 ns under pressure $p$ = 0.1 GPa (**c**), 0.08 GPa (**d**) and 0.06 GPa (**e**). Green, yellow and red spheres represent Zr, S and O atoms, respectively.

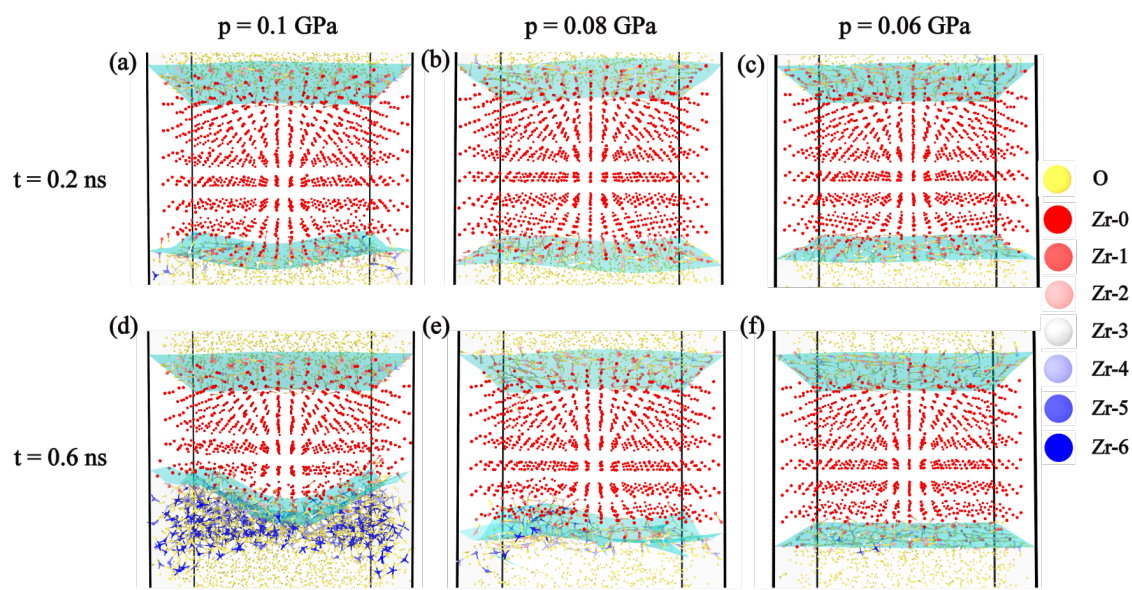

**Figure 2.** Snapshots of RMD simulations of $ZrS_2$ oxidation at different pressures and times at temperature 300 K. (**a-c**) $ZrS_2$ oxidation after time $t$ = 0.2 ns at pressure 0.1 GPa (**a**), 0.08 GPa (**b**) and 0.06 GPa (**c**). (**d-f**) $ZrS_2$ oxidation after time $t$ = 0.6 ns at pressure 0.1 GPa (**d**), 0.08 GPa (**e**) and 0.06 GPa (**f**). Oxygen atoms are colored yellow. Zr atoms are colored according to the number of coordinated oxygen atoms, *i.e.*, Zr atoms with no coordinated oxygen atoms are colored red, while the color of Zr atoms coordinated to more oxygen atoms is graded toward blue. To better visualize the oxide growth, surfaces of the oxidation front are shown in cyan color, and S atoms are omitted for clarity.

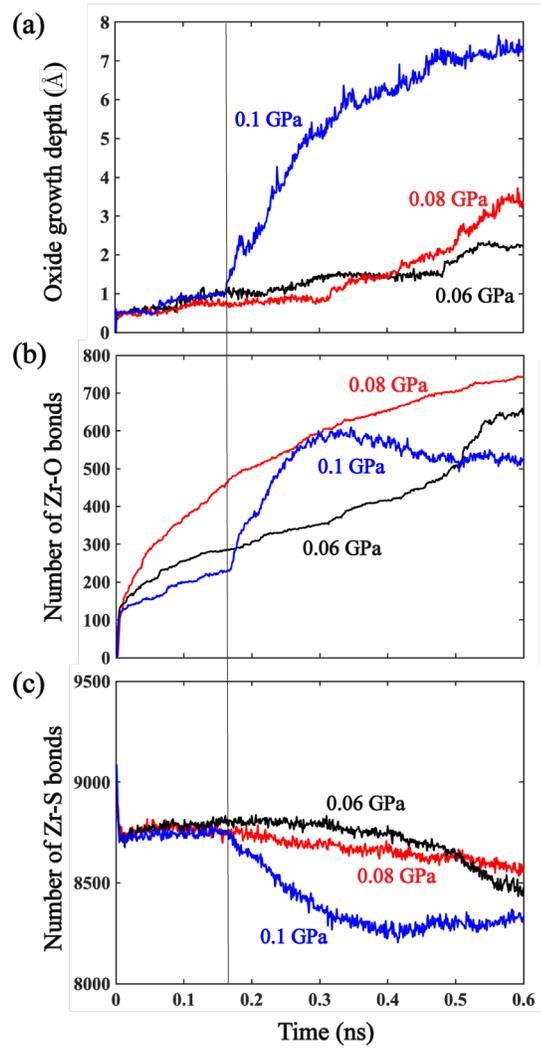

**Figure 3.** Pressure dependence of ZrS$_2$ oxidation dynamics. Time evolution of the oxide depth (**a**), number of Zr-O bonds (**b**), and number of Zr-S bonds (**c**). Blue, red and black curves represent oxidation at pressure 0.1, 0.08 and 0.06 GPa, respectively.

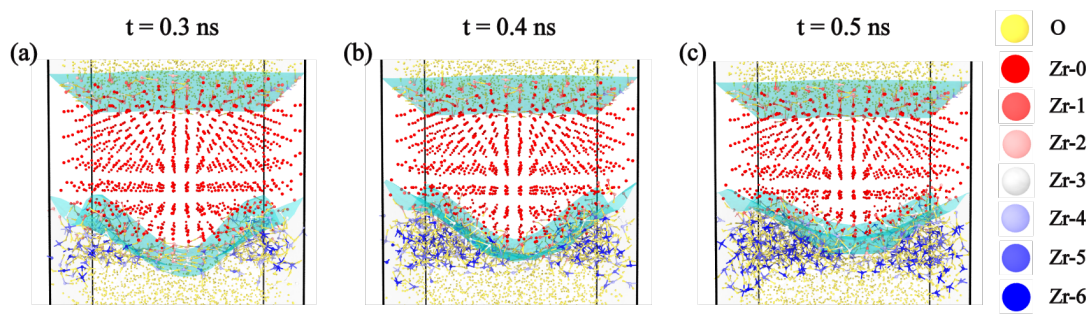

**Figure 4.** Snapshots of ZrS$_2$ oxidation under 0.1 GPa at time $t$ = 0.3 ns (a), 0.4 ns (b) and 0.5 ns (c). The color scheme is the same as that in Figure 2.

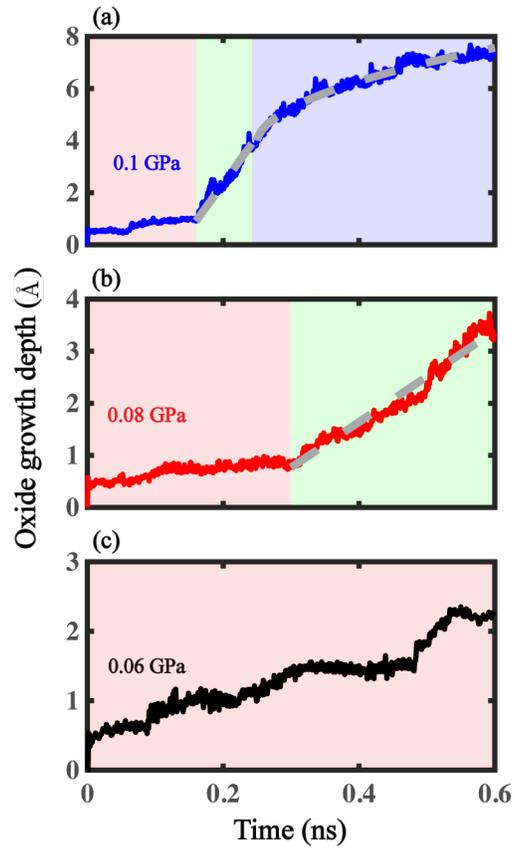

**Figure 5.** Fitting of the oxide growth depth *vs.* time for pressure (a) 0.1 GPa, (b) 0.08 GPa and (c) 0.06 GPa. Simulation results are shown blue, red and black, respectively, for 0.1, 0.08 and 0.06 GPa, while fits are shown as gray dashed lines. All fittings are with 95% confidence bound. The three stages of oxidation (I, IIa and IIb) mentioned in the main text are indicated by magenta, green and cyan backgrounds, respectively.

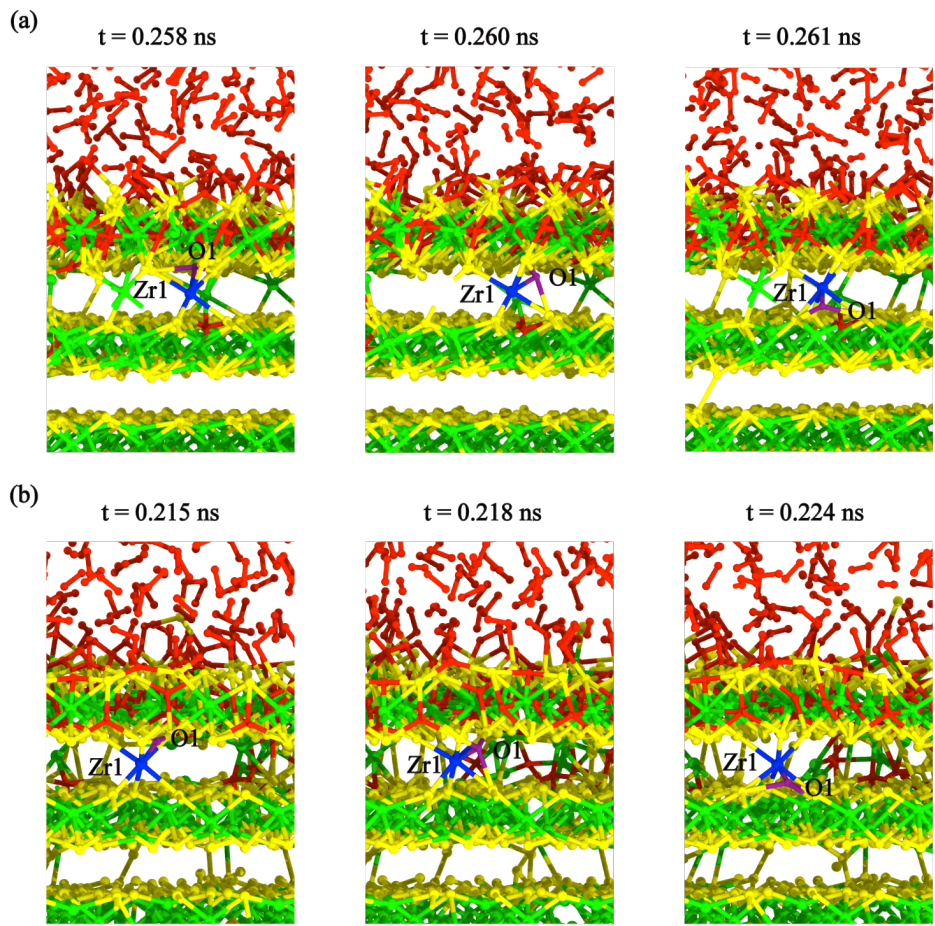

**Figure 6.** Snapshots of RMD simulation trajectory for ZrS$_2$ oxidation under 0.1 GPa. Two examples in (**a**) and (**b**) show the transport of O atoms (labeled O1 in purple color) assisted by the Zr atoms (labeled Zr1 in blue color) bonded to inner layer.

## ASSOCIATED CONTENT

**Supporting Information**

Supplementary Information (pdf format). Chemical Potential of Oxygen, Pair Distribution Function for Zr-O, Local Oxide Structures, Diffusive Oxygen Transport, Effects of Ensemble.

Supplementary Movies 1-3 (mp4 format). Oxidation dynamics under pressure 0.1 GPa (S1.mp4), 0.08 GPa (S2.mp4), 0.06 GPa (S3.mp4). Oxygen atoms are colored yellow. Zr atoms are colored according to the number of coordinated oxygen atoms, i.e., Zr atoms with no coordinated oxygen atoms are colored red, while the color of Zr atoms coordinated to more oxygen atoms is graded toward blue. To better visualize the oxide growth, surfaces of the oxidation front are shown in cyan color, and sulfur atoms are omitted for clarity.

**Author Contributions**

R.J. and L.Y. conceived the problem, and L.Y., R.K., A.N. and P.V. designed the simulations. L.Y. performed density functional theory and reactive molecular dynamics simulations. All authors discussed and prepared the manuscript.

**Funding Sources**

This work was supported as part of the Computational Materials Sciences Program funded by the U.S. Department of Energy, Office of Science, Basic Energy Sciences, under Award Number DE-SC0014607 (R.K., A.N., P.V.).

**Competing interests**

Authors declare that they have no competing interests.

**Data and materials availability**

All data are available in the main text or the supplementary information.

ACKNOWLEDGMENT

Simulations were performed at the Argonne Leadership Computing Facility under the DOE INCITE and Aurora Early Science programs and at the Center for Advanced Research Computing of the University of Southern California. This work was supported as part of the Computational Materials Sciences Program funded by the U.S. Department of Energy, Office of Science, Basic Energy Sciences, under Award Number DE-SC0014607 (R.K., A.N., P.V.).

# Supplementary Information
# for
## Pressure-Dependent Layer-by-Layer Oxidation of $ZrS_2$(001) Surface

Liqiu Yang, Rafael Jaramillo, Rajiv K. Kalia, Aiichiro Nakano, and Priya Vashishta

Correspondence to: anakano@usc.edu

## 1. Chemical Potential of Oxygen

Assuming an ideal gas behavior of $N$ oxygen molecules at pressure $p$ and temperature $T$, the chemical potential is obtained as

$$\mu = \left(\frac{\partial G}{\partial N}\right)_{T,p} \tag{S1}$$

through the Gibbs free energy,

$$dG = \left(\frac{\partial G}{\partial T}\right)_{p,N} dT + \left(\frac{\partial G}{\partial p}\right)_{T,N} dp + \left(\frac{\partial G}{\partial N}\right)_{T,p} dN = -SdT + Vdp + \mu dN. \tag{S2}$$

From the ideal gas equation of state, $pV = NkT$, and thus

$$\left(\frac{\partial G}{\partial p}\right)_{T,N} = V = \frac{NkT}{p}. \tag{S3}$$

When only varying pressure, we have

$$G(T,p) - G(T,p^0) = \int_{p^0}^{p} \left(\frac{\partial G}{\partial p}\right)_{T,N} dp = NkT \ln \frac{p}{p^0}, \tag{S4}$$

and thus

$$\mu_O(p,T) = \frac{1}{2}\mu_{O_2}(p,T) = \frac{1}{2}\left[E_{O_2} + \tilde{\mu}_{O_2}(p^0,T) + k_B T \ln \frac{p_{O_2}}{p^0}\right], \tag{S5}$$

where $p^0 = 1$ atm. Taking a reference state of $\mu_O(p,T)$ to be the total energy of oxygen in an isolated molecule [1], $E_{O_2} = \tilde{\mu}_{O_2}(T = 0\text{ K}, p) = 0$,

$$\mu_O(p,T) = \frac{1}{2}\left[\tilde{\mu}_{O_2}(p^0,T) + k_B T \ln \frac{p_{O_2}}{p^0}\right] = \tilde{\mu}_O(p^0,T) + \frac{1}{2}k_B T \ln \frac{p}{p^0}. \tag{S6}$$

Since the surrounding $O_2$ atmosphere forms an ideal-gas-like reservoir [2],

$$\tilde{\mu}_O(p^0,T) = \tilde{\mu}_O^{O-rich}(p^0, T = 0K) + \frac{1}{2}\Delta G(\Delta T, p^0, O_2)$$
$$= \frac{1}{2}[H(T,p^0,O_2) - H(0\text{ K},p^0,O_2)] - \frac{1}{2}T[S(T,p^0,O_2) - S(0\text{ K},p^0,O_2)]. \tag{S7}$$

$\tilde{\mu}_{O_2}(p^0,T)$ can be obtained using the enthalpy and entropy in thermochemical tables at $p^0 = 1$ atm (see the $\tilde{\mu}_O(p^0,T)$ values in Table S1) [2,3].

To characterize the pressure dependence of the chemical potential, we introduce $p^*$, for which $\mu_O(p^*,T) = 0$. Table S1 shows the calculated $p^*$ values at various temperatures. The calculated values are consistent with the numbers given in Soon *et al.* [1]. At temperatures 300, 600 and 900 K, they reported the pressure values to be $10^9$, $10^{10}$ and $10^{11}$ atm, respectively. Note from Table S1, $p^* = e^{20.888} = 1.179 \times 10^9 \cong 10^9$ atm at 300 K, $e^{23.596} = 1.768 \times 10^{10} \cong 10^{10}$ atm at 600 K, and $e^{25.272} = 0.945 \times 10^{11} \cong 10^{11}$ atm at 900 K.

## 2. Pair Distribution Function for Zr-O

To understand local structural changes associated with the oxidation dynamics in Figure 4, we have calculated the partial pair distribution function for Zr-O pairs (Figure S1). At time $t = 0.1$ and 0.2 ns, the first peak is located at 1.75 Å indicative of Zr-O chemical bond at all pressures (0.06, 0.08 and 0.1 GPa). However, at $t = 0.3$-0.6 ns, pair distribution exhibits large correlation at distance larger than 2 Å under the highest pressure of 0.1 GPa, along with the lowering of the first peak at 1.75 Å. This suggests structural disorder in the oxide, which is likely caused by fast incorporation of oxygen in $ZrS_2$ slab by mechanical forces, without sufficient time to form a well-defined chemical-bond network. The structural disorder under 0.1 GPa reconciles the decrease in Zr-O bond number in Figure 4b with the increase in oxide growth depth in Figure 4a compared to those at lower pressures.

## 3. Local Oxide Structures

To examine local oxide structures, we show the interfacial structure at time 0.6 ns under pressure 0.1 GPa in Figure S2. It shows locally disordered structures with Zr atoms bonding to different number of O atoms at different distances from the oxidation front. We have also calculated Zr-O bond-length distributions (Figure S3) and O-Zr-O bond-angle distributions (Figure S4) in the oxide layer at 0.6 ns under different pressures. For calculating O-Zr-O bond-angle distribution, we have used the Zr-O bond-length cutoff of 2 Å to be consistent with Zr-O bonds in $ZrO_2$ [4, 5]. We found that the Zr-O peak under 0.1 GPa is broader than those under lower pressures (0.08 and 0.06 GPa), which substantiate the structural disordering due to high pressure mentioned in the main text. This is likely due to the over-coordinated Zr and more Zr-O bonds with decreased bond order in the transient amorphous oxide. The broad bond-angle distributions in Figure S4 also indicate disordered oxide structures in the early stage of oxidation.

## 4. Diffusive Oxygen Transport

Despite the high reactivity at the oxide front under highest pressure (0.1 GPa), oxygen is still transported by diffusion through continuous amorphous oxide and driven by oxygen-concentration gradient. According to Fick's law, oxygen flux through the oxide layer is $F_O = -Ddc/dz$, where $D$ is the diffusion coefficient, $c$ is the oxygen concentration, and $z$ is the coordinate along the slab-normal direction. To demonstrate the diffusive behavior, we have plotted the mean square displacement (MSD) of oxygen atoms in the amorphous oxide after the onset of continuous oxide growth (0.16 ns) as shown in Figure S5. The MSD exhibits a diffusive behavior, from which the diffusion coefficient is estimated as $D = \langle|\Delta \vec{r}|^2\rangle/6t = 4.1 \times 10^{-4}$ (cm$^2$/s), which is consistent with the fitted $B$ parameter in the Deal-Grove model in the main text. The other ingredient of Fick's law is oxygen-concentration gradient, $dc/dz$. Figure S6 shows the oxygen-concentration profile in the oxide layer between the $O_2$ gas/oxide interface (left) and oxide/$ZrS_2$ interface (right) at time 0.47 ns. The oxygen concentration was calculated by dividing the oxide layer into cells of side 4.5 Å. The figure indeed exhibits linear drop of the oxygen concentration in the oxide layer. Together, Figures S5 and S6 support the applicability of the Deal-Grove model to the continuous oxide-growth stage as stated in the main text. Finally, it is worth noting that even the highest pressure of 0.1 GPa is orders-of-magnitude smaller than the elastic constant of $ZrS_2$ in the surface normal direction, $C_{33} \sim 30$ GPa [6]. As a result, we have not observed mechanical narrowing of the van der Waals (vdW) gap.

## 5. Effects of Ensemble

To examine the effects of ensemble on the oxidation behavior, we have performed additional reactive molecular dynamics (RMD) simulations in the isothermal-isobaric (NPT) ensemble instead of those in the canonical (NVT) ensemble used in the main text. The NPT simulations exhibit the same sequence of oxidation events as in the original NVT simulations, including the crucial gap-closing mechanism. Figure S8 compares snapshots of NVT and NPT simulations at a pressure of 0.1 GPa, showing similar gap-closing events.

**Table S1. Chemical potential of oxygen**

| $T$ (K) | 0 | 100 | 200 | 300 | 400 | 500 | 600 | 700 | 800 | 900 | 1000 |
|---|---|---|---|---|---|---|---|---|---|---|---|
| $\tilde{\mu}_O(p^0, T)$ (eV) | 0 | -0.08 | -0.17 | -0.27 | -0.38 | -0.50 | -0.61 | -0.73 | -0.85 | -0.98 | -1.10 |
| $k_B T$ (eV) | / | 0.0086 | 0.0172 | 0.0258 | 0.0345 | 0.0431 | 0.0517 | 0.0603 | 0.0689 | 0.0775 | 0.0861 |
| $\ln(p^*[\text{atm}])$ | / | 18.568 | 19.727 | 20.888 | 22.049 | 23.209 | 23.596 | 24.204 | 24.659 | 25.272 | 25.530 |

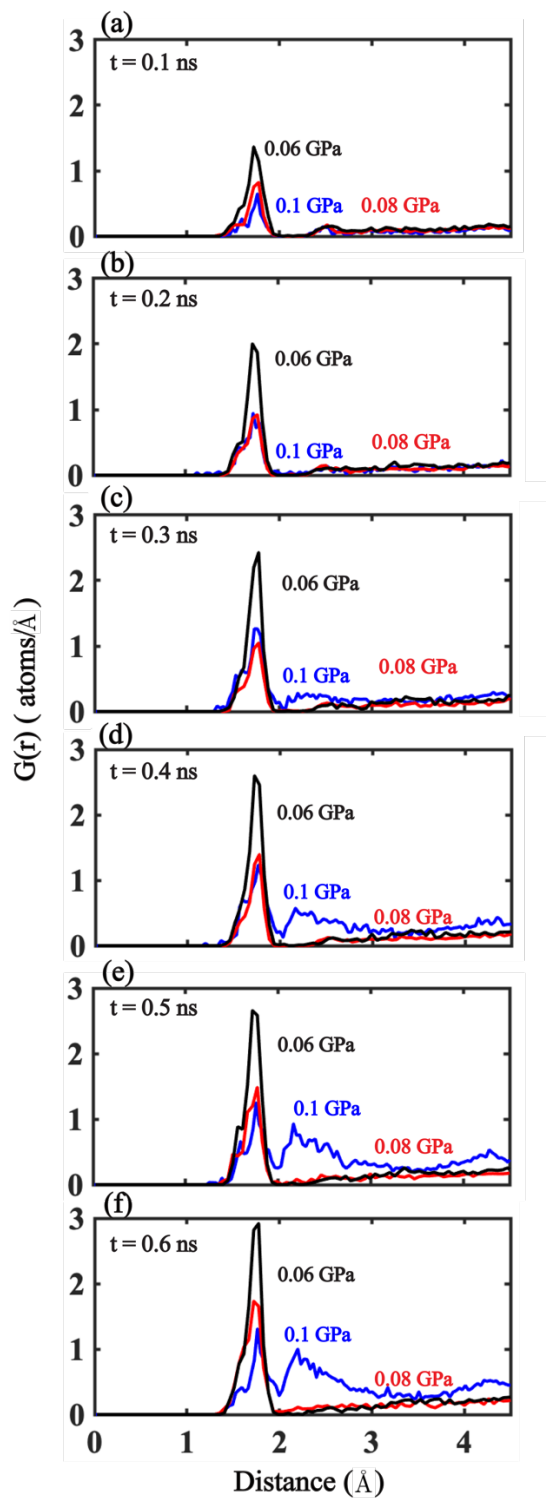

**Figure S1. Time evolution of Zr-O pair distribution function.** Blue, red and black curves are for pressure 0.1, 0.08 and 0.06 GPa, respectively, at time $t$ = 0.1 ns (**a**), 0.2 ns (**b**), 0.3 ns (**c**), 0.4 ns (**d**), 0.5 ns (**e**), and 0.6 ns (**f**).

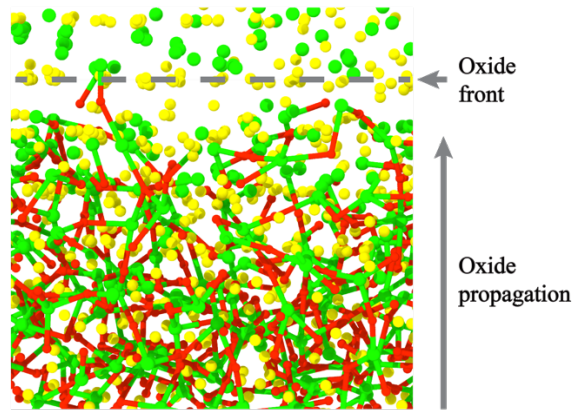

**Figure S2.** Snapshot of interface structure at time 0.6 ns under pressure 0.1 GPa. The color scheme is the same as that in Figure 1 in the main text. Zr-O bonds are drawn to highlight the progress of oxidation.

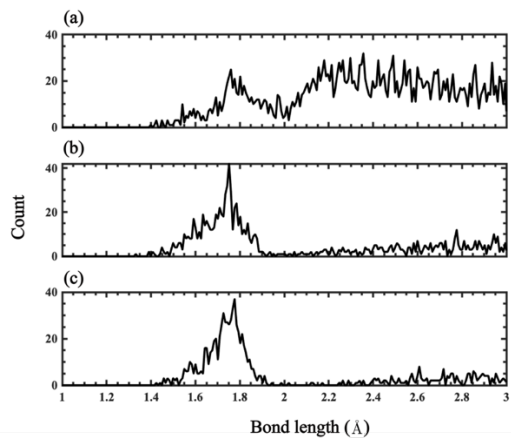

**Figure S3.** Zr-O bond-length distribution at 0.6 ns under pressure (a) 0.1 GPa, (b) 0.08 GPa, and (c) 0.06 GPa.

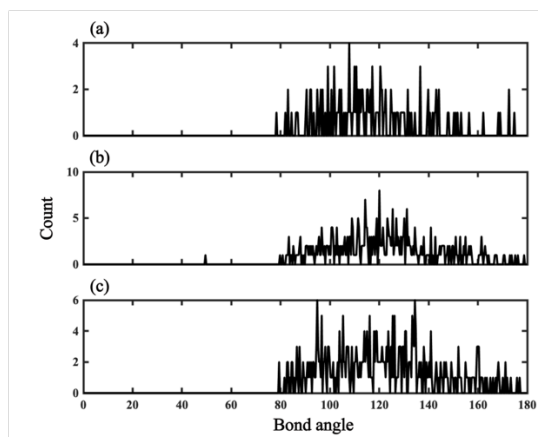

**Figure S4.** O-Zr-O bond-angle distribution at 0.6 ns under pressure (a) 0.1 GPa, (b) 0.08 GPa, and (c) 0.06 GPa.

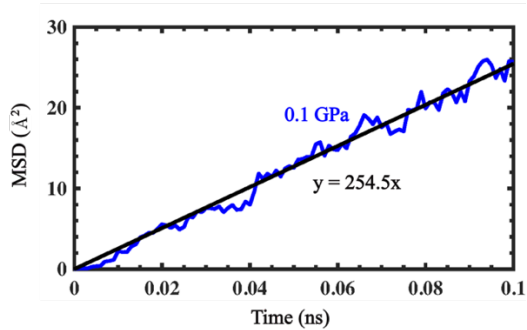

**Figure S5.** Mean square displacement of oxygen atoms in the amorphous oxide after the onset of continuous oxidation at time 0.16 ns under pressure 0.1 GPa.

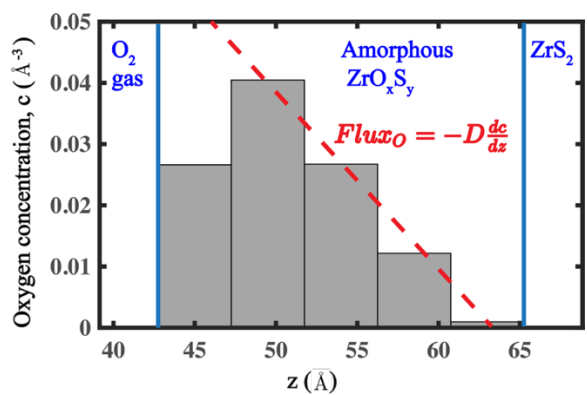

**Figure S6.** Oxygen concentration profile along the slab-normal direction, $z$, at time 0.47 ns under pressure 0.1 GPa.

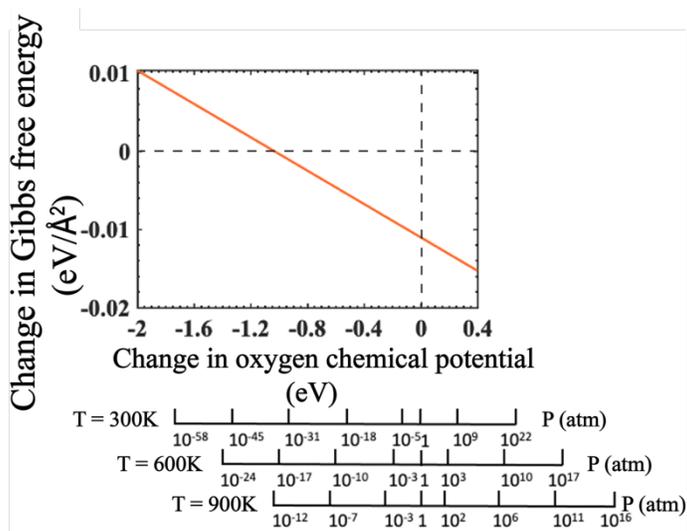

**Figure S7. Pressure dependence of the change of Gibbs free energy upon oxygen-atom adsorption.** Calculated change of Gibbs free energy, $\Delta G$, upon oxygen adsorption on ZrS$_2$(001) surface with varying oxygen chemical potential $\Delta\mu_O$. The corresponding pressure at various temperatures is shown as bars below the plot.

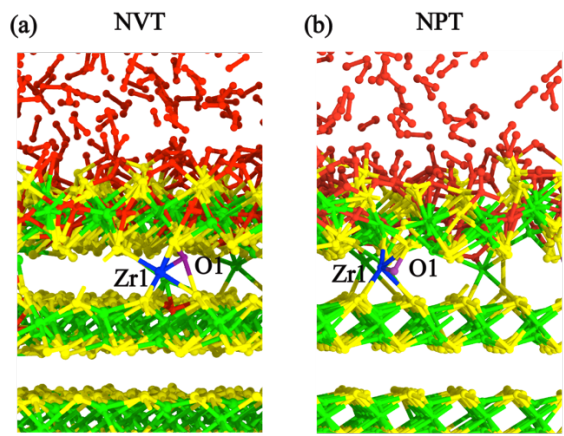

**Figure S8.** Snapshots showing O transport through gap closing for $ZrS_2$ oxidation under pressure 0.1 GPa in (a) NVT and (b) NPT ensembles. The color scheme is the same as that in Fig. 6 in the main text.

**Supplementary Movie 1 (mp4 format)**

Oxidation dynamics under pressure 0.1 GPa. Oxygen atoms are colored yellow. Zr atoms are colored according to the number of coordinated oxygen atoms, *i.e.*, Zr atoms with no coordinated oxygen atoms are colored red, while the color of Zr atoms coordinated to more oxygen atoms is graded toward blue. To better visualize the oxide growth, surfaces of the oxidation front are shown in cyan color, and sulfur atoms are omitted for clarity.

**Supplementary Movie 2 (mp4 format)**

Oxidation dynamics under pressure 0.08 GPa. The color scheme is the same as Supplementary Movie 1.

**Supplementary Movie 3 (mp4 format)**

Oxidation dynamics under pressure 0.06 GPa. The color scheme is the same as Supplementary Movie 1.